\documentclass[preprint,showpacs,preprintnumbers,amsmath,amssymb]{revtex4}
\usepackage{graphicx}
\usepackage{dcolumn}
\usepackage{bm}
\usepackage{epsfig}
\usepackage{subfigure}

\newcommand{\bi}{\begin{itemize}}
\newcommand{\ei}{\end{itemize}}
\newcommand{\be}{\begin{eqnarray}}
\newcommand{\ee}{\end{eqnarray}}
\newcommand{\bbmatrix}{\left( \begin{array}}
\newcommand{\eematrix}{\end{array} \right)}

\begin{document}

\title{Theoretical Description of Pseudocubic Manganites}
\author{Chungwei Lin and Andrew.J.Millis}
\affiliation{ Department of Physics, Columbia University \\
538W 120th St NY, NY 10027}

\begin{abstract}
A comprehensive theoretical model for the bulk manganite system La$_{1-x}$(Ca,Sr)$_x$MnO$_3$ is presented. The model includes
local and cooperative Jahn-Teller distortions and the on-site Coulomb and exchange interaction. The model is 
is solved in the single-site dynamical mean field approximation using a solver based on the semiclassical approximation. 
The model semi-quantitatively reproduces the observed phase diagram for the doping $0 \leq x<0.5$ and implies that
the manganites are in the strong coupling region but close to Mott insulator/metal phase boundary. The results
establish a formalism for use in a broader range of calculations, for example on heterostructures.
 
\end{abstract}

\maketitle


\section{Introduction}
\begin{figure}[htbp]
\centering
   \epsfig{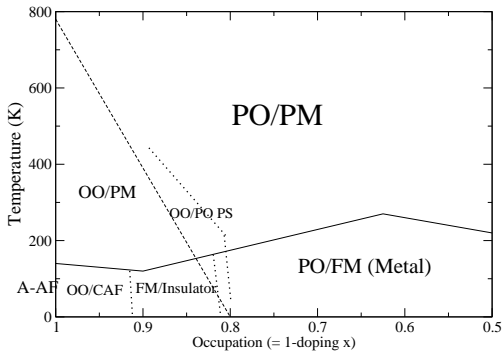}
   \caption{The experimental phase diagram as a function of doping $x$ and temperature $T$.
	PO: paraorbital; PM: paramagnetic; FM: ferromagnetic; OO: orbitally ordered; AAF: A-type
	antiferromagnetic. See text for the descriptions of the phases.
	}
   \label{fig:PhaseDiagram}
\end{figure}

LaMnO$_3$ crystallizes in a structure closely related the basic ABO$_3$ perovskite form. As the temperature is varied
it undergoes orbital ordering and antiferromagnetic transitions. Replacing some of the La by divalent
alkali ions such as Ca yields an even wider range of phenomena, including charge ordering, ferromagnetism, and
colossal magnetoresistance \cite{Goodenough_55, Wollan_55, Schiffer_95}. The phase diagram is summarized in Fig(\ref{fig:PhaseDiagram}).
While the manganites have been studied for many years, and much of the physics has been understood, there is as
yet no consensus in the literature on a model which is rich enough to account for all the physics, includes all of
important interactions, and can be solved to predict (or at least explicate) new phenomena such as those occurring in
systems such as heterostructures. In this paper we develop such a model and explore its properties.
Our results place the materials in the strong/intermediate coupling regime. 

\begin{figure}[htbp]
\centering
\epsfig{file = 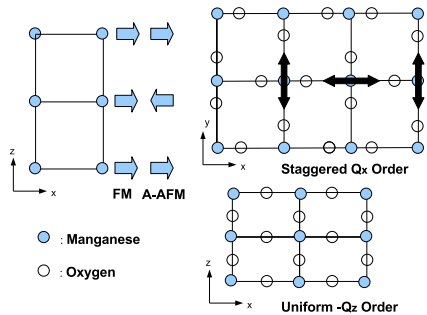,  width = 0.4\textwidth}
   \caption{(Color online) Illustrations of FM, A-AFM, and orbital orders. The filled and open circles represent manganese and 
	oxygens respectively. The light arrows stand for the core-spin orientation at each plane. Upper left panel:
	illustration of two magnetic orders. For FM, the spins 
	at different planes are aligned in the same direction while for A-AFM, the spins at adjacent planes arranged
	oppositely. Upper right panel: illustration of the in-plane staggered $Q_x$ order is shown in the up-right 
	corner. The double-arrows represent the long O-Mn-O distance caused by the $Q_x$ distortion. 
	Lower panel: illustration of the uniform $-Q_z$ order: 
	the system uniformly shrinks in the $z$ direction while it expands in $x-y$.
	}
   \label{fig:PhaseIllustration}
\end{figure}

The phase diagram shown in Fig(\ref{fig:PhaseDiagram}) includes
two magnetic orders, A-type antiferromagnetic (A-AFM) and ferromagnetic (FM) states, corresponding to the Mn spin
arrangements shown in Fig(\ref{fig:PhaseIllustration}).
The A-AFM structure consists of ferromagnetic planes antiferromagnetically coupled.
With our coordinate choice, each FM plane is spanned by $\hat{x}$ and $\hat{y}$ while the remaining direction is $\hat{z}$.
The orbital order (OO) in this context refers to a particular distortion arrangement where the oxygen octahedra have in-plane staggered 
($x-y$ plane) $Q_x$ JT distortions plus a uniform $-Q_z$ distortion (the minus sign
represents the octahedron shrinking in $z$ while expanding in $x-y$ directions) (Fig(\ref{fig:PhaseIllustration})). 
The metal/insulator phase boundary is determined from the DC resistivity.
The definition of metal/insulator is ambiguous. Here we adopt the definition that the system is
metallic/insulating at a given $temperature$ if the temperature derivative of resistivity is positive/negative.

In this paper, we present a model which captures all of the physics discussed above and solve it by the single-site 
dynamical mean field theory (DMFT)\cite{DMFT_96}. There are two main purposes for this study.
First, although basic understanding for exhibited phases at a $given$ doping is known, 
it is important to determine the extent to which the general model with {\bf a fixed set of parameters} matches the observed phase diagram.
Second, we wish to apply this theory to understand the behavior of the recently synthesized manganite 
superlattices \cite{Adamo_08, May_07}. 

Solving a theoretical model ordinarily requires approximations. Here we use the single site DMFT\cite{DMFT_96}.
This approximation requires as an intermediate step the solution of a quantum impurity model. In this paper we
solve the impurity model using a generalization of the semiclassical approximation (SCA) \cite{Okamoto_05}.
We generalize it to the 2-band case and develop a formalism for
incorporating the cooperative Jahn-Teller (JT) effect into the single-site DMFT. 
We semiquantitatively reproduce the observed phase diagram for $0<x<0.5$ and identify the sources of the observed phases. 
Our calculation yields three main results.
First, our calculation suggests the problem is in the ``strong/intermediate'' coupling regime in the sense that
under the single-site DMFT approximation the local interaction strength is slightly larger than the critical value
needed to drive a metal-insulator transition.
Consequently the system is very sensitive to the mechanisms governing the bandwidth such as magnetic order and details of crystal structure.
Second, the cooperative Jahn-Teller effect is the main source accounting for the observed high orbital ordering temperature. 
Finally, our calculation confirms that 
when the doping is increased to the colossal magneto-resistance
(CMR) region $x \sim 0.3$ ($N \sim 0.7$), the double-exchange (DE) mechanism becomes dominant. 

The rest of the paper is organized as follows. We first present the model and the interactions included. After
providing key steps for our approximation, we show how we fit parameters and then present the results.
Discussion concerning the inadequacies of our model/approximation and differences between the calculation and the experiments
is given in Section VII. Section VIII is a conclusion.
In the appendices we examine the validity of the SCA approximation and give details of the procedure 
we use to take the cooperative Jahn-Teller effect into account.

\section{Model Hamiltonian}
In this section we describe the interactions included in our model
and define terms and notations which shall be used for the rest of the paper.

{\em Tight-binding}: The band structure is described by a tight-binding model
where only nearest neighbor hopping between $e_g$ orbitals is included. 
A justification for this approximation is given in \cite{Ederer_07}. Two $e_g$
orbitals are labeled as $|1 \rangle = |3z^2-r^2 \rangle$, $|2 \rangle = |x^2-y^2 \rangle$.
This implies a band Hamiltonian which may be written as
\begin{eqnarray}
H_{band} 
= \sum_{\vec{k}, ab, \sigma} \epsilon_{\vec{k}, ab, \sigma} c^{\dagger}_{\vec{k}, a,\sigma}  c_{\vec{k},b,\sigma}
\label{eqn:H_band} 
\end{eqnarray}
$\epsilon_{\vec{k}, ab, \sigma} = -t (\varepsilon_0 \hat{e} + \varepsilon_z \hat{\tau}_z+ \varepsilon_x \hat{\tau}_x)_{ab}$
where $\hat{\tau}$ 's are Pauli matrices, $\hat{e}$ is the unit matrix and $\varepsilon_0 = \cos k_x + \cos k_y + \cos k_z $, 
$\varepsilon_z = \cos k_z -\frac{1}{2} (\cos k_x + \cos k_y )$, and
$\varepsilon_x = \frac{\sqrt{3}}{2} (\cos k_x - \cos k_y)$.
$a,b$ label orbitals, $i,j$ sites, and $\sigma$ spins. We emphasize that what denoted here as
two $e_g$ orbitals are actually the anti-bonding combination of Mn $3d$ and its neighboring 
oxygen $2p$($\sigma$ bond) states \cite{Ederer_07}. 

{\em On-site e-e}: For the on-site interaction within $e_g$ orbitals, we use the
Goodenough-Kanamori-Slater approximation in which the form of interaction is the same
as in the free atom. 
Two independent parameters conventionally denoted as $U$ and $J$ are required to specify this interaction.
It is generally accepted\cite{Hesper_97} that the charging energy $U$ may be
strongly renormalized by solid state effects whereas the inter-orbital exchange energy $J$ is less affected.
The e-e interaction within the $e_g$ multiplet is
\begin{eqnarray}
H_{EE} &=& \sum_{\sigma,\sigma'} (U- J) n_{1, \sigma} n_{2, \sigma'} + U \sum_{
i=1,2}n_{i,\uparrow} n_{i, \downarrow } + J( \, c^{\dagger}_{1,
\uparrow} c^{\dagger}_{1, \downarrow} c_{2, \downarrow} c_{2,
\uparrow} +h.c.)- 2 J \vec{S}_1 \cdot \vec{S}_2 
\label{eqn:H_EE}
\end{eqnarray}
where $\vec{S}_{1 (2)} = \vec{\sigma}_{\alpha \beta} c^{\dagger}_{1(2),\alpha} c_{1(2),\beta} $.
The $J( \,c^{\dagger}_{1, \uparrow} c^{\dagger}_{1, \downarrow} c_{2,\downarrow} c_{2, \uparrow} +h.c.)$ term is referred to as 
the pair hopping and the $- 2 J\vec{S}_1 \cdot \vec{S}_2$ term is the exchange.

{\em Hund's coupling}:
The coupling between Mn $e_g$ and Mn $t_{2g}$ electrons is approximated by
$H_{Hund}$ in which three $t_{2g}$ electrons are treated as an electrically inert ``core spin''
of magnitude $S=(3/2) \hbar$. We shall further approximate the core spin as classical and normalize $J_H$ by taking
$|\vec{S}|=1$, leading to 
\begin{eqnarray}
H_{Hund} &=& -J_H \sum_{i}\vec{S}_i \cdot c^{\dagger}_{i,\alpha} \vec{\sigma}_{\alpha \beta}
 c_{i,\beta} 
\label{eqn:H_Hunds} 
\end{eqnarray}
where $J_H>0$ and $|\vec{S}| = 1$.
The minus sign ensures that the high spin state is energy-favored in accordance with the Hund's rule.

{\em Lattice elastic energy}: 
\begin{figure}[htbp]
\begin{center}
   \epsfig{file = 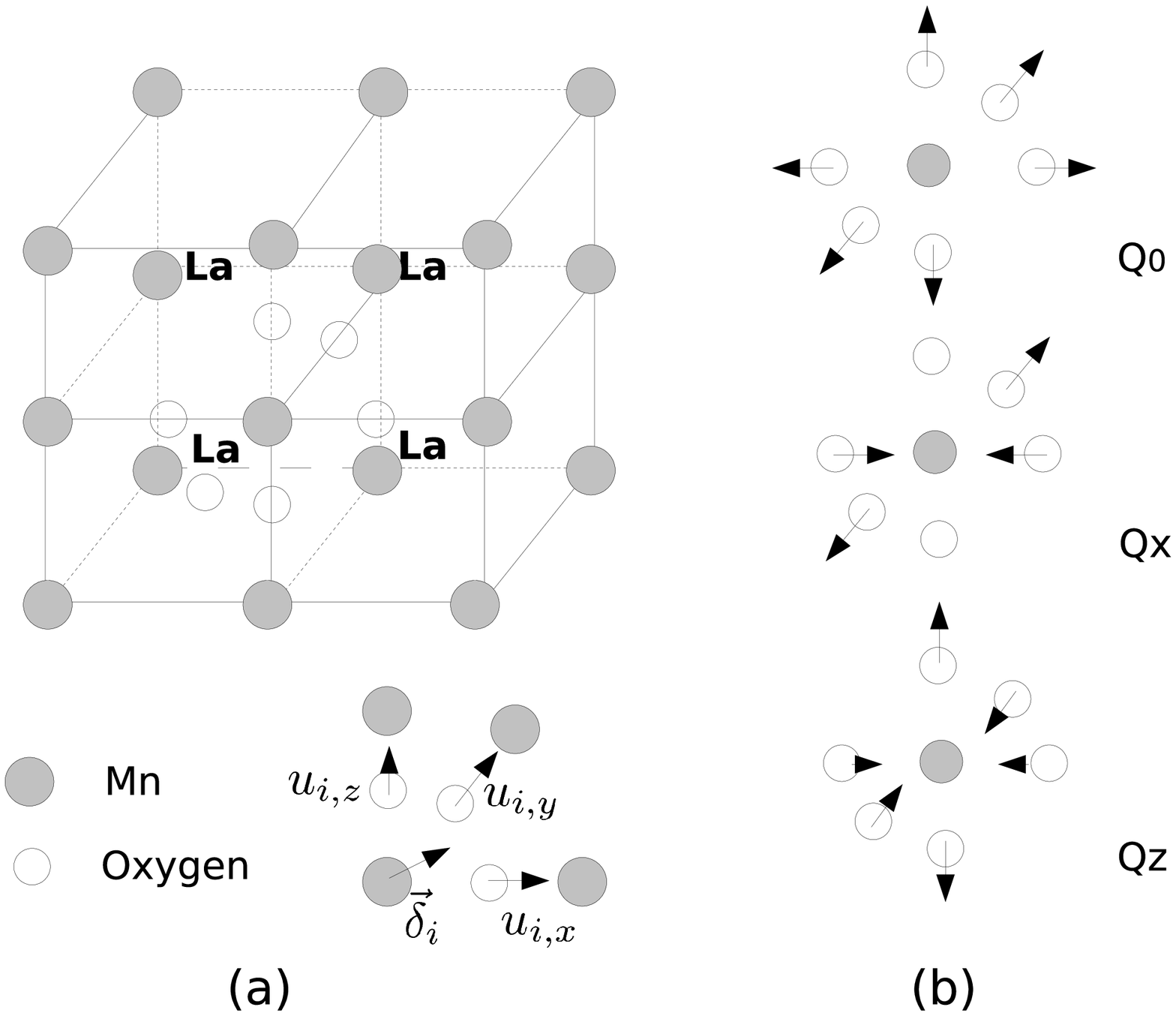, width=0.4\textwidth}
   \caption{(a) The ideal cubic perovskite structure for LaMnO$_3$ and the lattice degrees of freedom considered
	here: Mn can move in arbitrary direction $\vec{\delta}_i$ while oxygen ions only move along the Mn-O bond direction $u_{i,x(y,z)}$.
	(b) Sketch of the three octahedral distortion modes considered here: $Q_0$ breathing mode, $Q_x$ and $Q_z$ Jahn-Teller modes.}
   \label{fig:Cubic_JT}
\end{center}
\end{figure}
For the lattice degree of freedom, we consider Mn motions in arbitrary directions($\vec{\delta}_i$) and
oxygen only along Mn-O ($\sigma$) bond direction ($u_{i,x}$) \cite{Millis_96-2, Ahn_98} 
where $\vec{\delta}_i$ and $u_{i,x}$ are illustrated in Fig(\ref{fig:Cubic_JT}). 
The general lattice elastic energy in the harmonic approximation is
\begin{eqnarray}
H_{lat}&=& \frac{1}{2 K_{M-O} } \sum_{i,a} [(\delta^a_{i}-u^a_{i})^2 + (\delta^a_{i}-u^a_{i-a})^2] \\
&+& \frac{1}{2}\sum_{\vec{k},a b} E^{ab}(\vec{k}) \delta^a_{\vec{k}} \delta^b_{-\vec{k}} + 
\frac{1}{2}\sum_{\vec{k},a b} D^{ab}(\vec{k}) u^a_{\vec{k}} u^b_{-\vec{k}}
\label{eqn:H_lat}
\end{eqnarray}
where $1/K_{M-O}$ is spring constant between neighboring manganese, oxygen, while $E^{ab}(\vec{k})$ and $D^{ab}(\vec{k})$
are general Mn-Mn, O-O couplings in $k$ space. In the specific numerical calculations presented here, we set 
$E^{ab}(\vec{k})=0$ and $D^{ab}(\vec{k})=\frac{4}{k_{M-M}}\delta^{ab} \sin^2(k_a/2)$, but effects arising from
a more general interaction are discussed.
With our convention, $u$, $\vec{\delta}$ and lattice constants $K$ all have dimension of energy.

{\em Electron-Lattice coupling}:
The breathing ($Q_0$) and Jahn-Teller (JT) ($Q_x$, $Q_z$) modes at site $i$ are defined by:
\begin{eqnarray}
Q_{i,0} &=& \frac{1}{\sqrt{3}} (v_{i,x}+v_{i,y}+v_{i,z}) \nonumber \\
Q_{i,x} &=& \frac{1}{\sqrt{2}} (v_{i,x}-v_{i,y}) \nonumber \\
Q_{i,z} &=& \frac{1}{\sqrt{6}} (-v_{i,x}-v_{i,y}+2 v_{i,z}) 
\label{eqn:JTmodes}
\end{eqnarray}
where $v_{i,a} = u_{i,a}-u_{i-a,a}$. 
The $e_g$ orbitals couple to these three modes as
\begin{eqnarray}
H_{JT} &=& -\sum_{i,a,b}  ( Q_{i,x} \tau^x_{ab} + Q_{i,z} \tau^z_{ab})
 c^{\dagger}_{i,a} c_{i,b}  
\label{eqn:H_JT0} \\
H_{B} &=& - \beta Q_{i,0} (n_i - \langle n \rangle ) 
\label{eqn:H_B0}
\end{eqnarray}
where $\beta$ is dimensionless and positive. We will take $\beta = 1$ which simplifies the discussion of
cooperative Jahn-Teller effect. In this paper we treat $H_B$ by a mean field approximation, so it is only
important when the charge distribution is not uniform, like in the heterostructures
or in the charge-ordered phase. In our definition, positive $Q_z$ stands for the distortion
where the octahedron {\em expands} in the $z$ direction while $shrinking$ uniformly in $x-y$ with fixed volume.
The minus sign in Eq(\ref{eqn:H_JT0}) means that the positive $Q_z$ favors the occupancy of the $|3z^2-r^2 \rangle$ state.
This sign choice is justified because
positive $Q_z$ increases the lattice constant in $z$ direction and consequently reduces
$|t_{pd}|$ and $E_{anti-bonding}$, increasing the occupation in the anti-bonding band which is mainly composed of Mn $|3z^2-r^2\rangle$.
A similar consideration leads to the minus sign in Eq(\ref{eqn:H_B0}) (positive $Q_0$ means a volume expansion of the octahedron). 

{\em Cubic term in lattice energy}: An anharmonic cubic term \cite{Kanamori_61} in lattice energy is also included.
\begin{equation}
H_{Cubic} = -A (3 Q^3_{i,z}- Q^2_{i,x}Q_{i,z})
\end{equation}
where $A$ in our convention has the dimension [E]$^{-2}$.
Note this is the only cubic combination satisfying the lattice cubic symmetry. 
With the minus sign, positive $A$ is required to produce the observed distortions for LaMnO$_3$. 

{\em G-type AF coupling}: 
There is an isotropic nearest neighbor AF coupling (G-type) between $t_{2g}$ spins $\vec{S}_i$. 
\begin{eqnarray}
H_{AF} = J_{AF} \sum_{i,\hat{n}} \vec{S}_i \cdot \vec{S}_{i+\hat{n}} 
\end{eqnarray}
with positive $J_{AF}$. 
This coupling arises from the super-exchange mechanism (virtual hopping in $t_{2g}$ channels) and experimentally
shows in the G-type AF order exhibited in CaMnO$_3$ \cite{Wollan_55}.
The main effect of this term is to reduce the magnetic transition temperature.

The total Hamiltonian is then
\be
H_{tot} = H_{band}+H_{EE}+H_{Hund}+H_{lat}+H_{JT}+H_{B}+H_{Cubic}+H_{AF}
\ee

\section{Method}

We use the single-site Dynamical Mean Field Theory (DMFT) with the semiclassical approximation (SCA) to solve this two-orbital 
problem \cite{Okamoto_05}. In the DMFT approximation one replaces the full lattice self energy $\Sigma(\omega, \vec{p})$
by a local (momentum-independent) quantity $\Sigma(\omega)$ which is determined from the solution of an auxiliary
problem (quantum impurity model) plus a self consistency condition. The multiplicity of orbitals and interactions means
that the impurity model is not easy to solve. We use a Hubbard-Strotonovich transformation proposed by Sakai \cite{Sakai_04}
and the semiclassical approximation. To evaluate the frequency sum we use a procedure recently introduced by Monien \cite{Monien}.

We also mention two simplifications here. First we do not take into account the Coulomb potential produced by
the random distribution of cations  -- the only effect of replacing some La by divalent elements is
to reduce the $e_g$ electron population. However due to the screening effect from conduction electrons, we believe
this simplification is not crucial. Second, we restrict our calculation to charge-uniform states therefore we cannot obtain
the charge ordered phase which may be energy favored around half doping.

The rest of this section is organized as follows. Two key ingredients will be discussed: 
first we show how we encode the cooperative Jahn-Teller effect in the local impurity problem; second we give
some detailed formalism about SCA in this 2-orbital problem, especially how we decompose the
quartic interaction and what simplifications we have made. Then we discuss what measurements we used to fit parameters.

\subsection{Cooperative Jahn-Teller}
The local octahedral distortions ($Q_0, Q_x, Q_z$) at different sites are not independent --
distortion at one site inevitably causes distortion at the neighboring sites so that some global configurations of the lattice 
distortions are energy favored. This is
the cooperative Jahn-Teller effect \cite{Millis_96-2} which correlates the octahedral distortions at different sites.
Here we include this inter-site effect into the single-site DMFT by integrating out all of distortion fields except 
for those involving the variable $v$ at the site of interest.
The detailed calculation is given in the appendix A and the resulting local effective potential is
\begin{equation}
V_{eff}(Q_0, Q_x, Q_z) = \frac{Q^2}{2 K} + \epsilon \vec{F} \cdot \vec{Q}
\label{eqn:Coop_JT}
\end{equation}
where $K$ is an effective spring constant, $\epsilon \vec{F}$ represents the force exerted on the distortions at one site
by static (mean field) distortions on the other site. Here $\vec{F}$ measures the amplitude of the long ranged order and
$\epsilon$ gives the strength of the cooperative Jahn-Teller coupling. 

\subsection{On-site e-e}
The key step in our solution of the impurity model is to rewrite the quartic interaction into sums of complete squares so the
continuous Hubbard-Strotonovich transformations can be applied. Using the decomposition proposed
by Sakai \cite{Sakai_04}, we define
$f_{\sigma} \equiv c^{\dagger}_{1 \sigma} c_{2 \sigma} + c^{\dagger}_{2 \sigma} c_{1\sigma}$,
$n \equiv n_1+n_2$, $q \equiv n_1-n_2$, $s \equiv (n_{1,\uparrow}-n_{1,\downarrow}) + (n_{2,\uparrow}
-n_{2,\downarrow})$, and $d \equiv (n_{1,\uparrow}-n_{1,\downarrow}) - (n_{2,\uparrow}
-n_{2,\downarrow})$, and re-express Eq (\ref{eqn:H_EE}) as
\begin{eqnarray}
H_{EE} = U_0 n - \frac{J}{2} (f_{\uparrow} - f_{\downarrow})^2
+\frac{U_n}{2} n^2 - \frac{U_q}{2} q^2 - \frac{U_s}{2} s^2 -
\frac{U_d}{2} d^2 \label{eqn:H_square}
\end{eqnarray}
with $U_0=0$, $U_n=(3U-5J)/4$, $U_q=(U-7J)/4$, $U_s=(U+J)/4$, and $U_d=(U-3J)/4$. 
Due to the fermionic identity $\hat{n}_{i,\sigma}^2 = \hat{n}_{i,\sigma}$ ($i=1,2$ $\sigma=\uparrow,\downarrow$),
those coefficients are not unique. For example, $U_0=J/2$, $U_n=(3U-6J)/4$, $U_q=(U-6J)/4$, $U_s=(U+2J)/4$, and $U_d=(U-2J)/4$
is another legitimate set of choice. If the impurity problem is solved exactly, these two choices lead to 
the same result, but if approximate methods are used this needs not to be the case. However in the current study, 
the coefficients will be determined by fitting to data so this ambiguity is not important. 

\subsection{The Impurity Problem}
The impurity problem is then described by the effective action
$S=S_0+S_{int}$ where
\begin{eqnarray}
S_0 = -\int d \tau d \tau' a_{\alpha \beta}^{ i j}(\tau-\tau')
c^{\dagger}_{\alpha, i} (\tau) c_{\beta, j} (\tau')
\label{eqn:S_0_tau}
\end{eqnarray}
and $S_{int} = \int d \tau H_{EE} (\tau)$. The partition function is
$Z_{imp} = \int d\vec{Q}\, d\vec{S}\, D[c^{\dagger} c] \, e^{-S}$. 
Applying the Hubbard-Stratonovich (HS) transformations \cite{Sakai_04} to decouple $H_{EE}$, one arrives
\begin{eqnarray}
S_{int} &=& \int d\tau \left( \frac{1}{2 U_n} \phi^2_n(\tau) +
\frac{1}{2 J} \phi^2_f(\tau) + \frac{1}{2 U_q} \phi^2_q(\tau) +
\frac{1}{2 U_s} \phi^2_s(\tau) + \frac{1}{2 U_d} \phi^2_d(\tau)
\right) \nonumber \\ &+& \int d\tau \left( 
- i  \phi_n(\tau)  n(\tau) + \phi_f(\tau) (f_{\uparrow}(\tau) - f_{\downarrow}(\tau) ) 
 + \left[ \phi_q(\tau) q(\tau)+
\phi_s(\tau) s(\tau) +\phi_d(\tau) d(\tau) \right] \right) \nonumber \\
&+& \int d\tau \left(  \vec{Q} \cdot \vec{T}_{ab}
\delta_{\alpha \beta} +  J_H \vec{S} \cdot \delta_{ab}
\vec{\sigma}_{\alpha \beta} \right) c^{\dagger}_{i, \alpha} (\tau)
c_{j, \beta} (\tau)
\end{eqnarray}
To maintain the symmetries of the local interaction (SU(2) for spin, U(1) for orbital), we generalize the scalars 
$\phi_s$, $\phi_q$, to vectors $\vec{\phi}_s$ ($ = (\phi_{s,x}, \phi_{s,y}, \phi_{s,z})$, 3 components) and 
$\vec{\phi}_q$ ($ = (\phi_{q,z}, \phi_{q,x})$, 2 components) and average over their directions\cite{Schulz_90}. 
After expressing the $S_{int}$ in frequency space, two simplifications are made.
First, only zero frequency component for each HS field ($\phi(i \omega_0) = \phi$) is kept and 
second, saddle point approximations are applied to $\phi_f$, $\phi_d$, and $Q_0$ fields, i.e. $\phi_f = \phi_d =Q_0= 0$.
Different methods have been proposed for handling the $i \phi_n$ field \cite{Okamoto_05, Millis_96}. 
For the 2-band model studied here, we found the method of Ref\cite{Okamoto_05} effectively enhances the local orbital moment
as the doping increases which is opposite the observation while the method in Ref\cite{Millis_96} is free from this trouble, therefore
we follow Ref\cite{Millis_96} and take $i \phi_n=0$.
After integrating out the fermionic degree of freedom and combining the lattice effect, one gets
\begin{eqnarray}
V_{eff} = \left( \frac{Q^2}{2 K}
 + \frac{\phi^2_q}{2 U_q}  + \frac{\phi^2_s}{2 U_s}  \right) +A (3 \langle Q_z \rangle^2 - \langle Q_x \rangle^2 ) Q_z +
\epsilon \vec{F} \cdot \vec{Q} - T \sum_{\omega_n} \mbox{Tr} \log {\bf A} (i\omega_n) 
\label{eqn:V_eff_ee}
\end{eqnarray}
with 
\be
{\bf A} = {\bf a} +(\vec{Q} + \vec{\phi_q}) \cdot \vec{\tau} + 
(J_H \vec{S} + \vec{\phi_s}) \cdot \vec{\sigma}
\ee
where ${\bf A}$ is a $4 \times 4 $ matrix and ${\bf a}$ the Weiss function. The $A (3 \langle Q_z \rangle^2 - \langle Q_x \rangle^2 ) Q_z$ 
term comes from the simple mean field approximation of the cubic lattice energy. 

\subsection{Parameters and Fitting}
The discussion above indicates that there are seven parameters to be determined: the hopping $t$, 
effective local JT coupling $U_Q$, effective magnetic coupling $U_s$, Hund's coupling $J_H$, core-spin AF coupling $J_{AF}$,
cooperative JT coefficient $\epsilon$, and anharmonic lattice energy $A$. The first six have
the dimension of energy (E) and will be measured in units of the hopping $t$ while the last one has dimension 1/E$^2$.

The hopping strength $t$ has been determined from the band structure calculation 
for the experimental observed structure of LaMnO$_3$ to be roughly 0.5eV ($\sim$5000 K) \cite{Ederer_07}. 
By contrast, the value appropriate to the ideal perovskite structure is roughly 0.65eV. The difference
is due mainly to the effect of the GdFeO$_3$ rotation. For the series La$_{1-x}$Ca$_{x}$MnO$_3$
the distortion depends weakly on the doping $x$ (less than 10$\%$). For La$_{1-x}$Sr$_{x}$MnO$_3$
the rotation angle is more $x$-dependent (up to $30\%$). 
Therefore calculations in which $t$ is taken to be independent of $x$ may be appropriate for the Ca series but
are unlikely to be adequate for the Sr series. 
We focus here on the Ca series. We note that the value
$t=0.5$eV is in good agreement with the spectral weight inferred from the optical conductivity experiments 
on the ferromagnetic phase of La$_{0.7}$Ca$_{0.3}$MnO$_3$ \cite{Quijada_98}.
The $J_{AF}$ is estimated from the Neel temperature of CaMnO$_3$, roughly $110 K \sim 0.01$eV \cite{Wollan_55}.
The $T_N$ of the Heisenberg model $H=\sum_{i,j} J_{AF} \frac{9}{4} \vec{S}_i \cdot \vec{S}_j$ with $|\vec{S}|=3/2$
obtained from simple mean field is $\frac{20}{9} J_{AF}$, from which we estimate $J_{AF} \sim 4.5$meV$\sim 0.009t$. 
The super-exchange argument (from virtual hopping of $t_{2g}$ electrons)\cite{Feiner_99} also leads to the same estimate. 
We further found that within the mean field approximation, $J_{AF}$ term only acts to reduce to magnetic transition
temperatures by 5-10$\%$ but does not result in any new magnetic order.
Including this term however substantially increases the calculation time, so we typically
set $J_{AF}=0$ to accelerate the converging processes.
The anharmonic lattice energy is taken to be $A=0.006t^{-2}$ so that our calculation reproduces the observed JT distortion around room 
temperature for LaMnO$_3$ \cite{Rodriguez_98}.

The remaining parameters $J_H$, $U_s$, $U_Q$, $\epsilon$ are fitted by comparing the calculated and 
observed optical conductivity in LaMnO$_3$. Generally the $\sigma(\omega)$ contains peaks  
corresponding to local excitation energies of the system.
If we treat the hopping $t$ as a small perturbation, then peaks in the optical conductivity are roughly
the energy differences between the excited states and the ground state of the {\em local} Hamiltonian. Since there
are 4 states (2 spins $\times$ 2 orbitals) at each site, we expect there are 3 main peaks in $\sigma(\omega)$
corresponding to three 2-electron final states. The
saddle point estimate from the local potential 
indicates these three peaks are located at $2U_Q(1+\epsilon)$ (correct spin, other orbital), 
$2(J_H + U_s)$ (same orbital, antiparallel spin), and $2U_Q(1+\epsilon)+2(J_H + U_s)$ (other orbital, antiparallel spin)
which essentially agrees with our calculated results shown in Fig(\ref{fig:N1_sigma}).
The issue is discussed further in Ref\cite{cLin_08-2}.
Experimentally there are two apparent peaks observed in LaMnO$_3$ \cite{Kovaleva_04} --
the lower one around 2eV ($\sim 4t$) while the higher around 4eV ($\sim 8t$). There are several minor structures
around 5-6eV which we do not consider. Fitting the two main peaks in optical data suggests 
$U_Q(1+\epsilon) \sim 4(t)$, $J_H+U_s \sim 8 (t)$. We choose $J_H=2.8$, $U_s=1.4$, $U_Q=2.1$, $\epsilon=0.05$.
We found that if $J_H+U_s$ is fixed the relative values of $J_H$ and $U_s$ do not change the result much as long as $U_Q>U_s$. If $U_Q<U_s$,
the orbital order is not stable against the magnetic order.
Since at $T=0$ the SCA reduces to the simple mean field approximation where expectation values are determined by their saddle point values,
the combination $U_Q (1 + \epsilon)$ uniquely determines the $T=0$ phase. However $\epsilon$ has more significant
effect on the non-zero temperature phase. We choose $\epsilon$ so as to produce the observed $T_{oo}$. 

{\em Standard parametrization}: Unless indicated otherwise the results we shall present later correspond to our standard parameters
$t=0.5$eV, $U_Q=2.1t$, $U_s=1.4t$, $J_H=2.8t$, $J_{AF}=0$, $\epsilon=0.05$, $A=0.006/t^2$. We remind the reader that this choice
of $t$ is only appropriate for Ca doped materials; in the Sr series the $t$ changes with doping. 
 All the temperature, frequency are measured in $t$; a simple conversion $0.5$eV$\sim 5500K$. 

\section{The Phase Diagram}
\begin{figure}[htbp]
   \epsfig{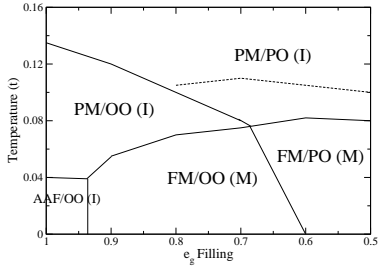}
   \caption{Calculated phase diagram as a function of doping $x$ and Temperature. M and I stand for metallic and
	insulating phases. The dashed curve is the PM/FM phase boundary, computed using $t=0.6$eV
	appropriate to La$_{0.6}$Sr$_{0.4}$MnO$_3$.
	}
   \label{fig:PhaseDiagram2}
\end{figure}

The calculated phase diagram as a function of doping and temperature is shown in Fig(\ref{fig:PhaseDiagram2}) 
The results qualitatively reproduce the observed phase diagram (Fig(\ref{fig:PhaseDiagram})) in the sense that
the relative positions of calculated magnetic/orbital phases are consistent with the experiments, but
the temperature scales are larger than observed.

According to the doping, we divide the phase diagram into three regions -- the undoped case ($x=0$, LaMnO$_3$),
the CMR regime ($0.3<x<0.5$), and the crossover regime ($0<x\leq0.3$). In essence, the undoped case is
cooperative JT dominated and the signature is the lattice distortions or equivalently the orbital order.
On the other hand, the CMR regime is double-exchange (DE) dominated in which the system is FM/metallic at low temperature.
In the crossover regime, both mechanisms play non-negligible roles to the system and we see that as doping increases,
the cooperative JT effect decreases ($T_{oo}$ decreases) while the DE mechanism gradually takes over ($T_c$ increases).
In principle, we can extend our calculation to $x>0.5$. However this region the effect of G-type AF coupling $J_{AF}$ starts
to emerge (or both double-exchange and cooperative JT effect decreases) and a different self-consistency condition (G-AF)
is required, so we shall leave it for future study.


The remainder of the paper is organized as follows.
We shall devote one section for the undoped case and one for the CMR and crossover region for more detailed discussions, then 
discuss the discrepancies between calculated and observed results. 
What important physics we are missing in our model/approximation and their effects to the current results
will be stated. We also point out here that for the spectral functions we shall present, the fermi energy is at zero. 
Without further indications, $\rho$, $\sigma$ and $A$ stand for resistivity, conductivity, and spectral function respectively.


\section{The Undoped Case}
\subsection{Overview}

Experimentally LaMnO$_3$ is insulating for all temperatures at least up to 800K \cite{Tobe_01, Okimoto_97, Kovaleva_04} 
, which is slightly greater than the orbital ordering temperature $T_{oo}$. 
When the temperature is lowered, it first goes from PO/PM to OO/PM at $T_{oo} \sim 780 K (0.135t)$, then from
OO/PM to OO/AFM at $T_{AAF} \sim 140K (0.04t)$ where temperatures in Kelvin are experimentally derived while
numbers in parentheses are from calculation. 
After a discussion about the nature of the insulating behavior, we examine the physical origins of the exhibited
phases.



\subsection{The Transport and Excitation Spectrum}
\begin{figure}[htbp]
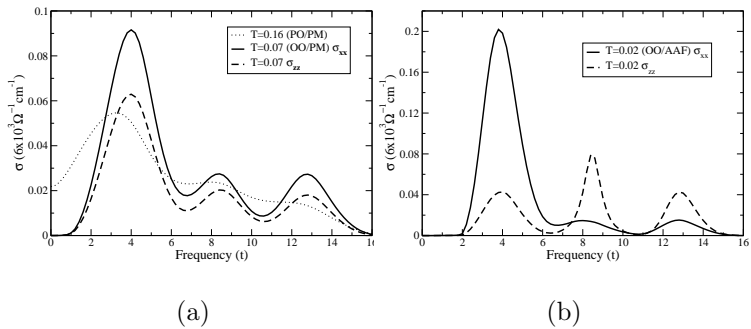

\centering
   \subfigure[]{\epsfig{file = Cond_PiPi0_PerMn_t1.0_Cubic0.006_JAF0_eps0.05_UQ2.1_JH2.8_Us1.4_Occu1.0_Temp0.18-0.07.eps, 
	width=0.3\textwidth}}
   \subfigure[]{\epsfig{file = Cond_PiPi0_PerMn_t1.0_Cubic0.006_JAF0_eps0.05_UQ2.1_JH2.8_Us1.4_Occu1.0_Temp0.07-0.02.eps, 
	width=0.3\textwidth}}
   \caption{ Optical conductivities for (a) $T=0.16t (> T_{oo})$ (dotted) $T=0.07 \sim 0.5 T_{oo}$, OO/PM phase 
	and (b) $T=0.02 t \sim 0.5 T_{AAF}$, OO/AAF phase.
	The heavy solid/dashed curves represent the in-plane/out-of-plane optical optical conductivities. 
	To convert the frequency scale to physical units we note that the band theory indicates $t=0.5$eV so
	$t=4$ corresponds to 2eV.
	}
   \label{fig:N1_sigma}
\end{figure}

In this subsection we present our calculated optical conductivities for the stoichiometric end-member LaMnO$_3$. We show that
the calculated conductivities are in good agreement with experiment and that the agreement implies that LaMnO$_3$ is 
a Mott insulator; a more careful discussion regard to Mott insulator is given in Ref\cite{cLin_08-2}. 

To establish our claim we present in Fig(\ref{fig:N1_sigma}) 
the optical conductivities for electric fields parallel to the $x-y$ plane (solid line) and perpendicular to it (dashed line)
at temperatures $T=0.16t$ (greater than $T_{oo}$), $T=0.07t$ (below $T_{oo}$, above the magnetic ordering temperature
$T_{AAF}$), and $T=0.02t$ (roughly $0.5 T_{AAF}$). 
The integrated optical conductivities upto 3eV qualitatively agree with experiments of LaMnO$_3$ \cite{Tobe_01}.
At $T=0.16t$ we see that the conductivity has two peaks at
$\omega \sim 4t$ and $8t$, and a soft gap at $\omega=0$. If we suppress the orbital order,
forcing PO/PM solution down to lower temperature, we find that the low 
frequency conductivity decreases \cite{cLin_08-2}. When the temperature is lowered to 
$T=0.07t$ where the orbital order is well established; 
the peak positions remain essentially unchanged. An anisotropy produced
by the orbital order appears and the gap at low frequency becomes sharper. 

As the temperature is further decreased into the A-type antiferromagnetic state the peaks
sharpen and the anisotropy becomes more pronounced with an increase in $\sigma_{xx}$ and a decrease in
$\sigma_{zz}$ for $\omega \sim 4t$ and the converse behavior in the $\omega \sim 8t$ regime. This qualitative 
behavior was used by authors of Refs \cite{AhnMillis_00} and \cite{Kovaleva_04} to identify the lower feature as 
the transition to the maximal spin, orbitally disfavored final state and the higher feature
as the transition to a lower spin, orbitally favored state. We make the same identification here and 
have adjusted the crucial parameters $U_Q$ and $U_s$ to place these peaks at the experimentally correct
energies. Referring now to Fig(\ref{fig:N1_sigma})(a) we see that for these parameters the correlations are
already strong enough to produce an insulating state in the absence of the long ranged order which is one
characteristic of Mott insulator, although the ``soft'' nature of the gap places the materials close to
the Mott insulator/Metal phase boundary.

\subsection{Origins of Exhibited Phases}
Along the temperature-descending direction, we summarize our understanding by the
following statements:

(1) The staggered $Q_x$ order is a consequence of the cooperative JT effect, i.e. a combined effect
from local JT interaction and lattice elastic energy.

(2) The energy difference between $(\pi,\pi,\pi)$ $Q_x$ and $(\pi,\pi,0)$ $Q_x$ orders is very small,
of the order of meV.

(3) The uniform $Q_z$ order is a consequence of the staggered $Q_x$ order, arising from the cubic term in lattice energy.

(4) The uniform $-(+)Q_z$ order reduces(enhances) the inter-layer AF coupling and decreases(increases) the Neel temperature.

To justify the first statement, we perform the calculation without cooperative Jahn-Teller effect
and obtain an orbital ordering temperature of $0.06t$ ($\sim 330$K) which is far too low compared to the observation. 
The cooperative Jahn-Teller coupling arising from
the corner-shared octahedra facilitates the staggered $Q_x$ order. The physical picture is
quite straightforward -- a $Q_x$ distortion on one site induces an $-Q_x$ distortion on neighboring sites
in the same $x-y$ plane. 

\begin{figure}[htbp]
\begin{center}
   \epsfig{file = 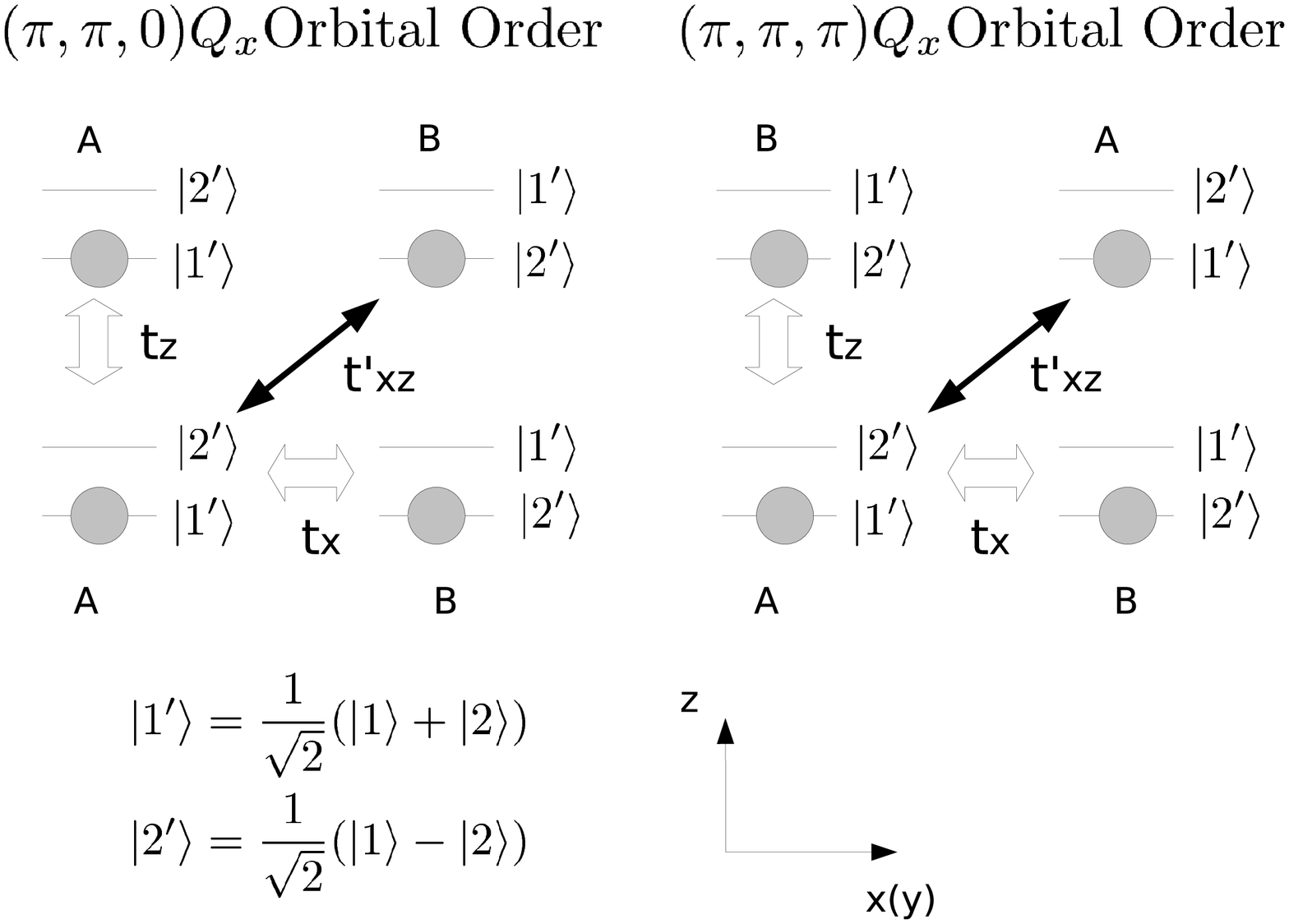, width=0.3\textwidth}%
   \caption{ Illustrations of superexchange processes in presence of different $(\pi,\pi,0)$ and $(\pi,\pi,\pi)$ $Q_x$ orbital order.
	A and B are two sublattices occupying local orbitals $(|1 \rangle + |2\rangle)/\sqrt{2}$ and
	$(|1 \rangle - |2\rangle)/\sqrt{2}$ respectively. The hopping matrices are directional and the explicit forms 
	are given in Ref\cite{Ederer_07}.
	}
   \label{fig:PiPi0_PiPiPiQx}
\end{center}
\end{figure}

The 2nd statement concerns the energy difference between the $(\pi,\pi,0)$ and the $(\pi,\pi,\pi)$ $Q_x$ order. 
There are two possible sources. The first one is the structure in the lattice contribution.
The simple form of the lattice Hamiltonian \cite{AhnMillis_01} we used in our numerical calculation has the same
restoring force for both distortions, but a more general form given in Eq(\ref{eqn:H_lat}) will distinguish them.
Assessing this possibility requires a DFT
calculation of phonon spectrum as is discussed in Ref\cite{AhnMillis_01}. The second possibility is the electronic 
energy which we now estimate from the super-exchange (SE, essentially 2nd order perturbation)
argument. The nearest neighbor ($t$) and second neighbor ($t'$) hopping processes which give rise
to superexchange are illustrated in Fig(\ref{fig:PiPi0_PiPiPiQx}).  

In terms of the local JT splitting of $\Delta_Q$, we find that the second order superexchange calculation yields that 
the energy gain for both orders are $-\frac{9}{4} \frac{t^2}{\Delta_Q}$. Therefore within SE approximation,
nearest neighbor hopping does not lift the degeneracy. However the second neighbor hopping does lift
the degeneracy. We find that the 
$(\pi,\pi,0)$ $Q_x$ state gains $-4 \frac{t'^2}{\Delta_Q}$ more energy than the $(\pi,\pi,\pi)$ state. From our DFT study \cite{Ederer_07},
$t' \sim 0.035$eV and $\Delta_Q \sim 1.4$eV, therefore the energy difference between these two orders is of the order meV
which is very small compared with other energy scales in the problem. We therefore believe that the lattice effect
is dominant. We model this by allowing only the $(\pi,\pi,0)$ order in our calculation.

\begin{figure}[htbp]
   \centering
   \epsfig{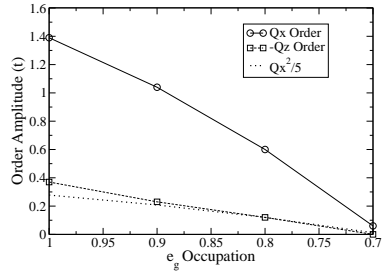}
   \caption{ The staggered $Q_x$ (solid curve) and the uniform $Q_z$ (dashed curve) orders as a function of doping at $T=0.1t$.
	In the bulk, the uniform $-Q_z$ order is induced by the cubic term in lattice energy and the magnitude
	is proportional to $Q_x^2$. For the chosen parameters here $Q_z \sim 0.2 Q_x^2$ (dotted curve).
	}
   \label{fig:Bulk_QxQz_T0.12}
\end{figure}

Within our approximation, the uniform $Q_z$ order is induced by the $local$ $Q_x$ distortion via
the cubic term\cite{Kanamori_61} in lattice energy, so the strength of the $Q_z$ order is 
proportional to $Q_x^2$. Fig(\ref{fig:Bulk_QxQz_T0.12}) shows the magnitudes of the staggered
$Q_x$ and uniform $Q_z$ orders at $T=0.1t$. For these parameters we found $Q_z \sim 0.2 Q_x^2$.

\begin{figure}[htbt]
\begin{center}
   \epsfig{file = 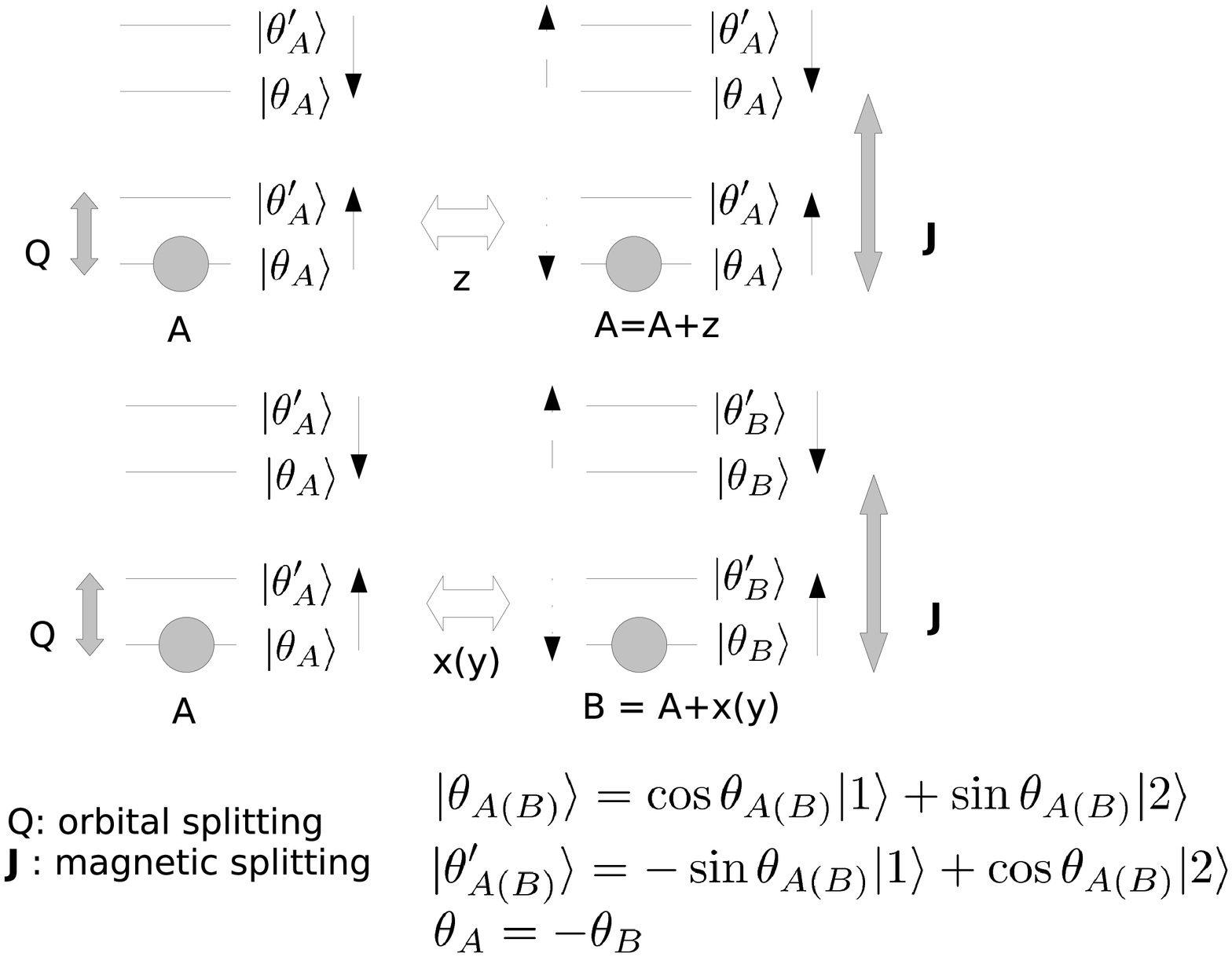, width=0.3\textwidth}
   \caption{Illustration of the $(\pi,\pi,0)$ $Q_x$ and a uniform $Q_z$ orbital order. The local orbital states
	are same in $z$ direction ($|\theta_A\rangle$ or $|\theta_B\rangle$) while are alternate between 
	$|\theta_A\rangle$ and $|\theta_B\rangle$ in the $x-y$ plane.
	}
   \label{fig:OrbitalOrder}
\end{center}
\end{figure}

\begin{figure}[htbp]
\centering
   \epsfig{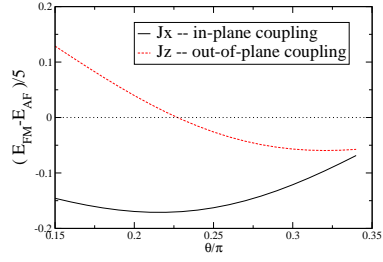}
   \caption{ (Color online) The out-of-plane ($J_z$) and in-plane ($J_x$) magnetic coupling estimated  
	from Eqn(\ref{eqn:Mcoupling}) with $\Delta_{JT}=4t$, $\Delta_{Hund}=8t$.
	The negative sign favors FM coupling. $\theta=0.25\pi$
	represents the case without uniform $Q_z$ order. Positive and negative $Q_z$ orders correspond to the region
	$\theta<\pi/4$ and $\theta>\pi/4$ respectively.
	}
   \label{fig:JxJz}
\end{figure}

The 4th statement concerns the relation between the magnetic order and orbital order. 
In particular it is the answer to the question that in the presence of a large staggered $Q_x$ order, how a small
uniform $Q_z$ order affects the magnetic order. We found that a small +/-$Q_z$ order can change the Neel temperature $T_{AAF}$
by as much as a factor of two. This effect can be qualitatively understood by comparing the effective magnetic couplings $J_i$ 
($i=z$ out-of-plane and $x$ in-plane) for different orbital orders using super-exchange arguments. 
The starting point is that for each site the electron occupies
the orbital $|\theta \rangle \equiv \cos\theta |1\rangle + \sin\theta |2\rangle$ ($0<\theta<\pi$)
which is the ground state of $-(Q_z \hat{\tau}_z + Q_x\hat{\tau}_x)$.
Since the strength of $Q_x$ order is at least 3 times larger than that of $Q_z$ order (see Section III G), 
we consider $Q_z/|Q_x|$ ranging from -0.3 to 0.3. 
As shown in Fig(\ref{fig:OrbitalOrder}) in the presence of the staggered $Q_x$ order, the system
is divided into two sublattices $A$ and $B$ on which the electron occupies orbital $|\theta_A\rangle$ and $|\theta_B\rangle$.
Defining $\cos 2\theta = Q_z/\sqrt{Q_x^2+Q_z^2}$, $\sin 2\theta = Q_x/\sqrt{Q_x^2+Q_z^2}$, one finds the
occupied orbitals at $A$ and $B$ are $|\theta\rangle$, $|-\theta\rangle$. Using the 2nd order perturbation, 
one estimates the magnetic couplings from the energy difference
between FM and AF spin configurations ($J=E_{FM}-E_{AF}$) as
\be
J_z (\theta) &=& -\frac{\cos^2 \theta \sin^2 \theta}{\Delta_{JT}} + \frac{\cos^4 \theta}{\Delta_{Hund}}
 + \frac{\cos^2 \theta \sin^2 \theta}{\Delta_{Hund}+\Delta_{JT}} \nonumber \\
J_x (\theta) &=& -\frac{3/16+\cos^2 \theta \sin^2 \theta}{\Delta_{JT}} + \frac{\cos^2 \theta - 3\sin^2 \theta}{4\Delta_{Hund}}
 + \frac{3/16+\cos^2 \theta \sin^2 \theta}{\Delta_{Hund}+\Delta_{JT}} 
\label{eqn:Mcoupling}
\ee
where $\Delta_{JT}$ and $\Delta_{Hund}$ are orbital and magnetic splitting respectively. From previous discussion we found
$\Delta_{JT}\sim2$eV$\sim4t$, $\Delta_{Hund} \sim 4$eV$\sim8t$. The corresponding results are given in 
Fig(\ref{fig:JxJz}) where $\theta/\pi=0.2, 0.25, 0.3$ correspond to $Q_z/|Q_x|=+0.3, 0, -0.3$ respectively. 
We see that for these $\theta$ values the in-plane magnetic coupling 
is always FM while the out-of-plane changes from FM to AF when $\theta/\pi$ varies from 0.3 to 0.2
(zero coupling at $\theta=0.22\pi$). This SE estimate therefore implies a positive $Q_z$ order 
is required to produce the observed AAF order. In our DMFT calculation, we always find the AAF order at low temperature,
but we indeed find the Neel temperature drastically (50$\%$) increases when we go from small $-Q_z$ to small $+Q_z$ order.
Thus the trend of variation of $T_N$ with strain is correctly captured by the superexchange calculation, but
other processes also contribute the overall sign.
This result indicates that the interaction is not strong enough to justify the superexchange approximation
but that the superexchange results does capture one aspect of the important physics.

\section{CMR Regime and Crossover}

\begin{figure}[htbp]
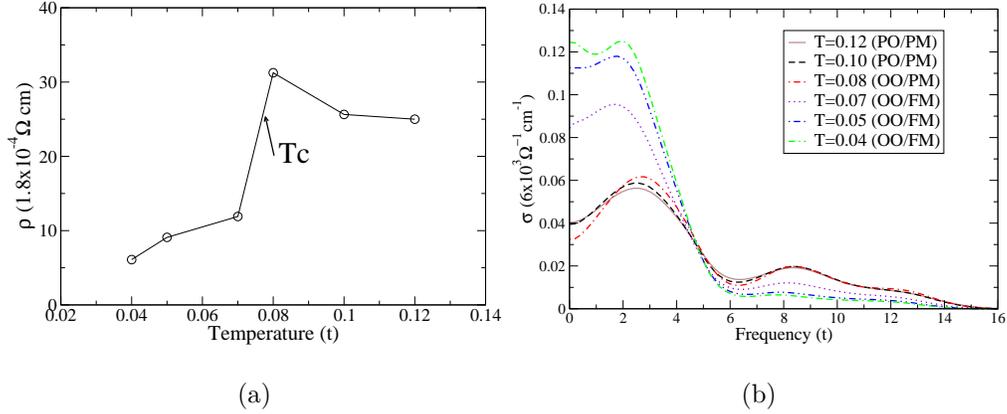

\centering
   \subfigure[]{\epsfig{file = sigma_xx_N0.7.eps, 
	width = 0.4\textwidth}}
   \subfigure[]{\epsfig{file = Cond_PiPi0_PerMn_t1.0_t1-0.00_Cubic0.006_JAF0._eps0.05_UQ2.1_JH2.8_Us1.4_Occu0.7_Temp0.1-0.04.eps, 
	width = 0.4\textwidth}}
   \caption{ (Color online) (a) In-plane DC resistivity $\rho_{xx}(T)$ as a function of temperature for $x=0.3$.
	(b) In-plane optical conductivities for $x=0.3$, $T=0.12 t$ to $0.04t$.
	}
   \label{fig:N0.7}
\end{figure}

\subsection{CMR Regime ($0.3<x<0.5$)}
We choose $x=0.3$ ($N=0.7$) as being representative of the CMR region. 
For this doping, the system goes from PO/PM to PO/FM around 275$K$. The transition is accompanied by an
insulator/metal transition which is shown in Fig(1) of Chapter 1 in Ref\cite{CMR}. At this transition the 
$d\rho/dT$ changes sign. Fig(\ref{fig:N0.7})(a) shows the calculated $\rho(T)$. We indeed find a M/I
transition across the Curie transition. This M/I transition is also reflected in the optical conductivities shown in 
Fig(\ref{fig:N0.7})(b). Our calculations qualitatively agree with 
the experimental data of La$_{0.7}$Ca$_{0.3}$MnO$_3$.  
To be more quantitative, Table I compares 
the kinetic energy defined as $K = (\frac{\hbar a}{e^2})\frac{2}{\pi} \int_0^{2.7eV} \sigma(\omega) d\omega$
obtained from both experiments \cite{Quijada_98} and our theoretical calculation. 
The reasonable agreement suggests that our model well captures
the main physics (right degree of freedom and reasonable effective interactions) below 2.7eV.
However there are several differences between calculation and data. First the experimental
values are systematically larger. Two reasons are that the experiments inevitably involve transition from lower oxygen bands 
to Fermi level which is not included in our model and our calculation yields orbital order at low T which does not occur in
the actual material. In addition our calculation 
overestimates the density of states around zero frequency at high temperature PM phase. In terms of
DC resistivity, it means the high-T PM phase is not insulating enough. We 
shall discuss the possible physics accounting for this inconsistency later.

\begin{center}
\begin{tabular}{|l||l|l|} \hline
     & Expt (Ref\cite{Quijada_98})  & Calculation \\ \hline
FM   & 0.22      & 0.152    ($T=0.04t$)   \\ \hline
PM   & 0.1       & 0.076    ($T=0.1t$)  \\ \hline
\end{tabular} \\ Table I: Kinetic energy in the unit of eV obtained from both
experiments and our calculation, using $t=0.5$eV.
\end{center}

One issue from earlier calculations
is that the $T_c$ for pure DE model is roughly 3 times higher than the observed one\cite{Michaelis_03}. 
The Curie temperature $T_c$ with $J_{AF}=0$
obtained here (roughly 0.08$t$) is $\sim 40\%$ lower than that of \cite{Michaelis_03}. 
Introducing G-type AF coupling $J_{AF}$($\sim 0.01t$ from our fit) further reduces $T_c$ to 0.075$t \sim 412K$,
not too far from the experimental value $\sim 275 K$. We expect that a large fraction of the remaining
difference arises from spatial and thermal fluctuation effects not captured by our mean field theory.



\subsection{The Crossover Regime ($0.1<x<0.3$)}
\begin{figure}[htbp]
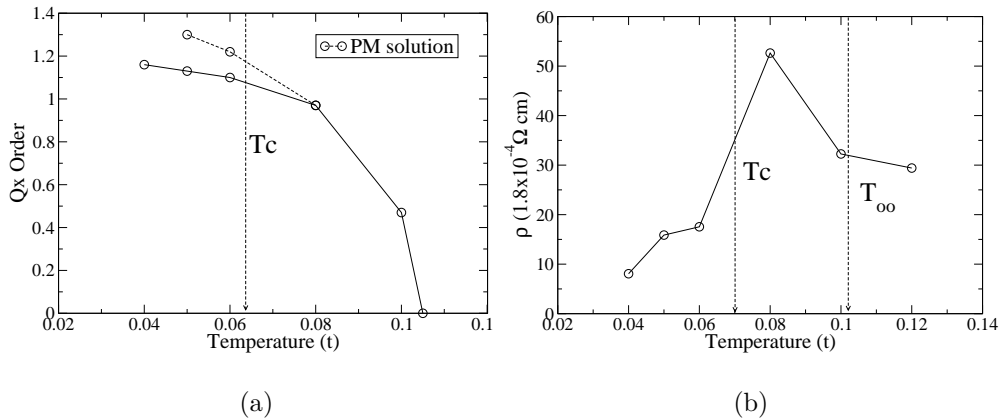

\centering
   \subfigure[]{\epsfig{file = Qx_N0.8.eps, width = 0.4 \textwidth}}
   \subfigure[]{\epsfig{file = sigma_xx_N0.8.eps, width = 0.4 \textwidth}}
   \caption{ (a) The amplitude of staggered $Q_x$ order as a function of temperature. The dashed curve are calculated
	at paramagnetic phase.  (b) Resistivity for $x=0.2$ ($N=0.8$) as a function of temperature.
	The vertical dashed lines indicate the transition temperatures.
	}
   \label{fig:N0.8}
\end{figure}

In this doping range, when the temperature is lowered, the system goes from PO/PM to OO/PM, then to OO/FM
phase and we take $x=0.2$ ($N=0.8$) as a representative doping. 
In this region, both cooperative JT and DE mechanisms are important. 
These two mechanisms are competing and not compatible in the following sense -- the cooperative JT tends to
break the in-plane symmetry which facilitates the staggered $Q_x$ order and localizes electrons, while 
the DE wants the system to be uniform and delocalizes electrons. This competition is shown in Fig(\ref{fig:N0.8})(a)
where the magnitude of the staggered $Q_x$ order is plotted. We find that when lowering the temperature, the magnitude of $Q_x$ order 
increases above the Curie temperature, and then quickly saturate below $T_c$. 
If we force the PM solution at low temperature, then the staggered $Q_x$ order keeps on increasing as T is decreased.
This is consistent with the pair distribution function (PDF)
measurements \cite{Billinge_00} which show below the Curie temperature at $x=0.25$, the peak associated with 
JT distortion decreases when lowering the temperature. Fig(\ref{fig:N0.8})(b) shows the resistivity as 
a function of temperature. We see that the system is an insulator at high temperature and a downturn in $\rho(T)$ happens at the Curie temperature, below which DE effects gradually takes over and the system is metallic.
Finally we point out that around $x=0.3$, $T_{oo}$ and $T_c$ happen around the same temperature (around $0.1t$ in
Fig(\ref{fig:PhaseDiagram2})). We do not resolve the behavior around this point carefully. 

\section{Discussion}
\subsection{Summary}
With a fixed set of parameters, our calculations semiquantitatively 
produce the observed phase diagram -- the relative positions of magnetic and orbital orders in the doping-temperature plane
are consistent with experiments. In particular the magnetic transition temperatures (both Neel and Curie temperatures)
are in reasonable agreement with data with calculated values being about 1.5 times higher than 
the measured values. Some part of the difference arises from the fluctuation corrections to the mean field theory,
which are typically of the order of $30\%$ in three dimension.
As for the excitation spectra, our results are consistent with observed optical conductivity.
In particular we reproduce the peak positions (this is how we fit some of the parameters) and the corresponding amplitudes
for a wide range of doping and temperature. 
We believe these agreements to experiments indicate that our model and fittings capture the essential physics of the manganite problem.
In this section we give a more detailed discussion on several issues and on inconsistencies to data
regarding to our results.

\subsection{Role and Effect of GdFeO$_3$ Rotation}
Our results indicate the local interaction strength is only slightly stronger than the critical value
of Mott transition \cite{cLin_08-2} implying the system is very sensitive the hopping $t$. 
As discussed in Ref\cite{Ederer_07,CMR}, the hopping is very sensitive to the structure. In particular the
manganites form in a distorted version of the ideal perovskite structure. The most important
important distortion appears to be a GdFeO$_3$-type rotation which buckles the Mn-O-Mn bond.
Table II summarizes the relation between the bond angle, the cation composition, and the hopping.
For a perovskite material AMnO$_3$, the Mn-O-Mn bond-angle as a function of
A-site composition is taken from Ref\cite{CMR} and the corresponding hoppings are calculated in Ref\cite{Ederer_07}. 
From this table we infer that using the same hopping $t=0.5$eV for LaMnO$_3$ and Ca doped manganite
is reasonable, but is not for Sr-doped. 
\begin{center}
\begin{tabular}{|l|l|l|} \hline
A-site     & bond-angle  & hopping (ratio)  \\ \hline
La$_1$ Ideal     & 180         & 0.65eV    (1)   \\ \hline
La$_{0.7}$Sr$_{0.3}$   & 166       & 0.58eV    (0.9)  \\ \hline
La$_{0.7}$Ca$_{0.3}$   & 160       & 0.53eV    (0.81)  \\ \hline
La$_1$ Real       & 155    & 0.5eV    (0.78)  \\ \hline
\end{tabular} \\ Table II: The composition of A-site elements, its corresponding
Mn-0-Mn angle and effective hopping $t$.
\end{center}

\begin{figure}[htbt]
\begin{center}
   \epsfig{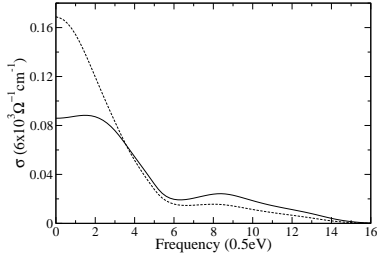}
   \caption{The optical conductivities calculated for parameters which simulate La$_{0.7}$Sr$_{0.3}$MnO$_3$.
	Solid and dashed curves are computed above and below the Curie temperature. To convert the frequency into
	physics units [eV], divide the x-axis by two.
	}
   \label{fig:t1.2_Cond}
\end{center}
\end{figure}

In Ref\cite{cLin_08-2}, we show that for $t=0.65$eV LaMnO$_3$ is not a Mott insulator. 
When using $t=0.6$eV to simulate La$_{0.7}$Sr$_{0.3}$MnO$_3$, we find that
(1) the Curie temperature increases from $\sim 420$K ($t=0.5$eV) to $\sim530$K (shown in the dashed curve
in Fig(\ref{fig:PhaseDiagram2})) and (2) the high temperature PO/PM phase becomes almost metallic ($d\rho/dT$ is very flat,
and the minimum around $\sigma(\omega=0)$ almost vanishes). 
The optical conductivities for $t=0.6$eV for two temperatures, just above and below $T_c$, are shown in
Fig(\ref{fig:t1.2_Cond}). Both of our findings
($T_c$ and $\sigma(\omega)$) are consistent with the difference between La$_{0.7}$Sr$_{0.3}$MnO$_3$
and La$_{0.7}$Ca$_{0.3}$MnO$_3$ reported in Ref\cite{Quijada_98}. We emphasize, however, that the main
message here is that for the given local interaction strength, the system is very sensitive to the bandwidth and
any uncertainty in estimating parameters could easily drive the system to either Mott insulating or metallic phases.

\subsection{Orbital Ordering}
We found that with our standard parameters, the calculated $T_{oo}$ decreases too slowly as a function of doping $x$ (see Fig(\ref{fig:PhaseDiagram2})). Fine tuning parameters (e.g. varying $\epsilon$ and $U_Q$)
can correct this problem but this degree of data fitting is somewhat arbitrary so we do not persue it here.

\begin{figure}[htbt]
\begin{center}
   \epsfig{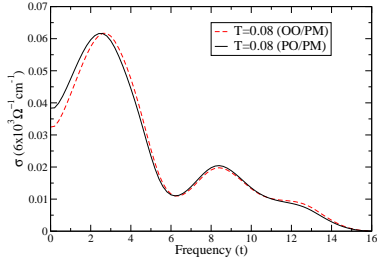}
   \caption{(Color online) The optical conductivities calculated at $x=0.3$, $T=0.08t$ for PO (solid) and OO (dashed) phases.
	}
   \label{fig:Cond_PO_OO}
\end{center}
\end{figure}

However, we point out that using our standard parameters,
the orbitally disordered and orbital ordered phases behave very similarly as far as the excitation is concerned. To
demonstrate this we show in Fig(\ref{fig:Cond_PO_OO}) the optical conductivities for $x=0.3$, $T=0.08t$ at both
OO and PO phases. One sees that the difference is very limited. Furthermore we found the Curie temperature obtained
from these two phases are very close (difference $<5\%$). These results suggest 
that the crucial electronic physics is controlled by local distortions which (because the
correlations are strong) are well formed.
The important effect caused by the orbital order is actually the uniform $Q_z$ order which introduces an isotropy to the system
and whose sign substantially affects the Neel temperature for the undoped case.

We also mention here that the effective Mn-Mn hopping being through oxygen $p$ orbitals also introduces an intersite orbital
coupling \cite{Mostovoy_04} which is very similar to the cooperative JT effect and is referred to as the
``charge-transfer'' mechanism. This can be understood from the 
super-exchange argument where we consider a simple Mn-O-Mn system and compare energies of different orbital
configurations by the perturbation expansion of the Mn-O hopping $t_{pd}$ \cite{Millis_97}. In the model
where the oxygen orbitals are not included, those ``virtual processes'' lead to an 
spin-independent orbital-exchange interaction \cite{Mostovoy_04} as
\be
H_{o-ex} = A \sum_{i, \alpha = x,y,z} I_i^{\alpha} I_{i+\alpha}^{\alpha} 
\ee
with $I^z = \tau_z$, $I^x = -\frac{1}{2} \tau_z - \frac{\sqrt{3}}{2} \tau_x$, $I^y = -\frac{1}{2} \tau_z + \frac{\sqrt{3}}{2} \tau_x$,
and $A$ a $positive$ coefficient. The simple mean field approximation to the term produces an external field on site o as 
\be
H_{o-ex} &=& A \sum_{\alpha = x,y,z} I_o^{\alpha} ( \langle I^{\alpha} \rangle_{+\alpha} + \langle I^{\alpha} \rangle_{-\alpha} )
\nonumber \\&=&
A \left[ \tau_z \frac{1}{\sqrt{6}}(2 E_z -E_x -E_y) + \tau_x \frac{1}{\sqrt{2}}( -E_x +E_y)  \right]
\ee
with $E_z = \frac{2}{\sqrt{6}} \sum_{\alpha = \pm z} \langle \tau_z\rangle_{\alpha}$ and
$E_{x(y)} = \frac{-1}{\sqrt{6}} \sum_{\alpha = \pm x (\pm y)} [\langle \tau_z\rangle_{\alpha} +(-) \sqrt{3} \langle \tau_x\rangle_{\alpha} ]$.
which is of the same form of the cooperative JT effect derived in the appendix. 
Therefore in our approximation where the orbital order and structural JT distortions are equivalent, 
including the charge-transfer mechanism amounts to a reinterpretation of our cooperative JT parameter $\epsilon = 2A$ and
does not change any results.

\subsection{High-T Insulating Phase}
Using the standard parameters, our calculation obtains an insulating behavior at high T PO/PM phase for
doping ranging from $x=0$ to $x\sim 0.4$. With the semiclassical approximation, the electron-electron interaction
is replaced by some classical fields and the impurity problem becomes polaron-like \cite{Okamoto_05, Millis_96}.
The high T insulating phase away from zero doping should be therefore interpreted as a phase separation
between $N=1$ orbitally fully-polarized state and $N=0$ state. Since our estimate indicates \cite{cLin_08-2}
the on-site Coulomb interaction is roughly three times stronger than the electron-lattice, the semiclassical
method may overestimate the insulating behavior under single-site DMFT approximation.

The other issue is that compared to the experiments, our calculations overestimate the optical conductivity around zero frequency at
high temperature PO/PM phase. This might be due to short-ranged correlations not included in 
the single site DMFT approximation.
According to recent cluster DMFT studies of the 1-band Hubbard model \cite{Park_08, WernerMillis_08}, 
including the short-range correlation significantly reduces the low energy density of states. We also observe that 
in doped CMR systems there is a strong
empirical association between insulating behavior ($d\rho/dT < 0$ with $\rho$ DC resistivity and $T$ temperature)
and strong short-ranged Jahn-Teller (polaron glass) order \cite{Lynn_07,Sen_07,Ward_08}.
Including spatial correlations beyond the single site approximation is an important topic for future regard.


\subsection{Missing Phases}
Our calculation misses two phases. 
First around $x=0.5$, a charge ordered (CO) phase occurs, 
accompanied by one particular orbital and magnetic order called CE phase \cite{Goodenough_55} which
requires a very large unit cell ($4 \times 4$) in the $x-y$ plane. Our in-plane unit cell
is not large enough to include this phase. However at $x=0.5$ we do find that the convergence becomes more and more
difficult when lowering the temperature (below $T=0.04t$) which may be an indication of CE phase. 
Second, around $x=0.1-0.2$ we do not get the FM insulating phase at low temperature.

\subsection{Limits of Approximation}
Now we discuss the limits of our approximation. 
First we discuss the breathing mode polaron effect.
In the current approximation the breathing mode coupling is treated in simple mean field and therefore
has no effect in the charge-uniform phase. To include the breathing mode polaron, one has to consider the
fluctuation of the breathing-mode distortion by integrating over $Q_0$ field when computing the 
impurity model \cite{Millis_96}. Since the real time-consuming computation involved in our approach is doing
multi-dimensional integral (see Section III.F), performing an additional integral is now beyond our computational power.
It is possible that the breathing mode polaron is also crucial for the charge order at $x=0.5$.
Since with breathing mode polarons electrons are already localized but just randomly distributed at high temperature
(therefore the system is not charge-uniform, but can be treated within single-site DMFT \cite{Millis_96}), 
the CO is then formed at low temperature to gain more energy from the gap. Without the polaron to
localize electrons, it is very hard to get CO (usually it requires some nesting in the band structure
which is not the case here). 
\begin{figure}[htbp]
   \epsfig{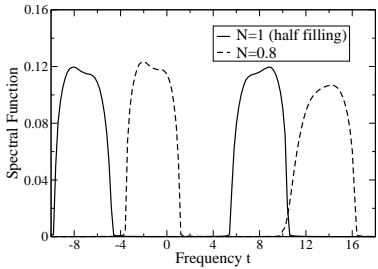}
   \caption{Spectral function calculated for one-band Hubbard model for bandwidth 6$t$, on-site $U=16t$ at $T=0.1t$.
	For both half filling (solid) and $N=0.8$ (dashed), the lower and upper bands have the equal weights.
	}
   \label{fig:Hubbard}
\end{figure}

Finally we point out that the SCA does not treat the quantum physics of the Mott insulator faithfully. 
To be specific, we take 1-band Hubbard model with strong coupling as an example.
With the SCA the metal/insulator transition occurs at $U\sim U_{c1}$ so the effect pf the Kondo peak is absent. Further,
independent of doping $x$, the upper and lower Hubbard bands have the same weight with SCA 
while in the reality, the upper Hubbard band represents adding one electron to the $occupied$ site the 
whose weight is ought to be 1-$x$. This consideration implies the SCA works well at half filling and becomes less reliable away from it. 
This is illustrated in Fig(\ref{fig:Hubbard}) which show the spectral functions for $N=1$ (half-filling) and $N=0.8$. 
We see that in both cases the upper and lower bands have the same weight. 
For our 2-band manganite model in the strong coupling limit, the SCA solution for PO/PM phase results in
4 bands with weights 1-$x$, 1+$x$, 1+$x$, 1-$x$ (from low energy to high) respectively. However the 3rd peak
corresponds to adding one electron to the state with same orbital but opposite spin whose weight should be 1-$x$. 
Based on the same argument we conclude that the SCA for 2-band model is more reliable without doping than with doping.
A more accurate treatment of the doped phase requires an improved, fully quantum impurity solver.

\section{Conclusion}
A general model for bulk manganite, including electron-electron, electron-phonon, and
phonon-phonon interactions is formulated and solved by semiclassical approximation.
Our calculation is qualitatively good in the sense that it yields the right distribution
of phases in the $(x,T)$ plane and produces the correct low energy excitations as described 
in section 2.5.2. The physical origin of each exhibited phase is identified within our model. 
For the LaMnO$_3$ below $T_{oo}$, the exhibited in-plane staggered $Q_x$ order is mostly driven 
by the cooperative Jahn-Teller (lattice effect) rather than the pure electronic effect while the 
uniform $-Q_z$ order is a consequence of the anharmonic term in lattice energy. 
Our results indicate that the local interaction strength is only slightly stronger than the critical
value for Mott transition and the system is consequently very sensitive to mechanisms controlling
the effective bandwidth. With this local interaction strength, the orbitally ordered and orbitally
disordered phases behave very similarly. 
As the doping increases, the electrons start to delocalize and after $x > 0.3$, the double-exchange 
mechanism dominates so orbital order vanishes and the system has the FM/Metallic ground state.

Discrepancies between our calculation and the observations are also carefully discussed. 
In particular our calculations overestimate the optical conductivity around zero frequency at
high-temperature insulating phase. This inconsistency leads us to conclude that the key physics we are missing 
in the calculation is the short-ranged correlation. In the future we will include the short-ranged correlation and also 
adopt a better impurity solver for this problem.

\section{Acknowledgment}
We thank Armin Comanac, Claude Ederer and Hartmut Monien for many helpful discussions, 
and DOE-ER46189 and the Columbia MRSEC for support.
\appendix
\section{The Validity of Semiclassical Approximation}
In this appendix we examine the validity of the semiclassical approximation by comparing the
excitation spectrum computed using the SCA results to the exact eigenstates of the local Hamiltonian. As discussed in Section II,
we assume that the crystal field (ligand field) is large enough that the $t_{2g}$ levels are in their
maximum spin state and that the pair hopping between $t_{2g}$ and $e_g$ orbitals is quenched. In this case
the on-site Hamiltonian in the $e_g$ manifold is 
\begin{eqnarray}
H_{loc} &=& \sum_{\sigma,\sigma'} (U- J) n_{1, \sigma} n_{2, \sigma'} + U \sum_{
i=1,2}n_{i,\uparrow} n_{i, \downarrow } + J( \, c^{\dagger}_{1,
\uparrow} c^{\dagger}_{1, \downarrow} c_{2, \downarrow} c_{2,
\uparrow} +h.c.)\nonumber \\ &-& 2 J \vec{s}_1 \cdot \vec{s}_2 
-2J_H \vec{S}_c \cdot (\vec{s}_1 + \vec{s}_2) + \Delta(n_1-n_2)
\label{eqn:H_loc}
\end{eqnarray}
Here $\vec{s}_i = \sum_{\alpha \beta} c^{\dagger}_{i \alpha} \vec{\sigma}_{\alpha \beta} c_{i \beta}$, $\vec{S}_c$
has magnitude $3/2$ and $\Delta$ is the crystal field splitting arising from the long-range Jahn-Teller order. 
In spherical symmetry $J_H=J$; we assume this henceforth. The eigenstates are characterized by the particle number,
total spin and total $e_g$ spin, and the orbital configuration. There are 16 1-electron and 24 2-electron eigenstates,
taking the configurations of the core spin into account. 

To compare the exact solution of the local Hamiltonian to experiment and the semiclassical calculation, we need the quantity
$\Delta E(S) = E(n=2, S)+E(n=0,S=3/2) - 2 E(n=1,S=2) $ which gives the locations of peaks
in the optical conductivity in the atomic limit. 
The following table lists the eigenstates and the corresponding transition energies.

\begin{center}
\begin{tabular}{|l|l|l|} \hline
States   & $\Delta E$  &  Semiclassical  \\ \hline \hline
$^3A_2 (5/2)$  (6) & $U-3J/2+2\Delta $        & $2U_Q$       \\
$^3A_2 (3/2)$  (4) & $U+7J/2+2\Delta $        & $2(U_s+J_H)+2U_Q$            \\
$^3A_2 (1/2)$  (2) & $U+13J/2+2\Delta $        & not accessible       \\ \hline
$^1E^- (3/2)$  (4) & $U+9J/2+2\Delta -\sqrt{4\Delta^2+J^2} $      & $2(U_s+J_H)$        \\
$^1A (3/2)$   (4) & $U+9J/2+2\Delta +\sqrt{4\Delta^2+J^2} $        & not accessible       \\
$^1E^+ (3/2)$  (4) & $U+7J/2+2\Delta $        & $2(U_s+J_H) + 2U_Q$       \\ \hline
\end{tabular} \\ Table III:
The 2-electron eigenstates and \\ the corresponding transition energies \\
\end{center}
Determining the coupling strength in Eq(\ref{eqn:H_loc}) by fitting the optical data \cite{Kovaleva_04}
is described in detail in Ref\cite{cLin_08-2}. Here we simply quote the results,
$U=2.3 \pm 0.3$eV, $2\Delta \sim J \sim 0.5$eV. Following the analysis and notations in Ref\cite{cLin_08-2},
there are three optical peaks located at
\be
\Delta E_{HS} &=& U-3J/2 + 2 \Delta \nonumber \\
\Delta E^{-}_{LS} &=& U+9J/2+ 2 \Delta  - \sqrt{4 \Delta^2 + J^2} \nonumber \\
\Delta E^{+}_{LS} &=& U+7J/2 + 2 \Delta \label{eqn:local_spectra}
\ee

We now compare this result to the semiclassical calculation. From Fig(\ref{fig:N1_sigma}) we observe three peaks in
the optical conductivity: a low-lying peak at energy $2U_Q$ which we identify with $\Delta E_{HS}$, an intermediate
peak at energy $U_s +J_H$ which we identify with $\Delta E^{\bar{JT}}_{LS}$, and a higher peak at the sum of these energies. 
This highest peak represents physically the states $^1E^+(3/2)$ and $^3A_2(3/2)$ where
both orbitals are occupied while the total spin (including the core spin) is 3/2. 
The $^3A(3/2)$ state in large $\Delta$ limit represents a state where both electrons occupy 
energy-disfavored orbital which cannot be reached by a single hopping and has no correspondence in the SCA. 
It is the defect of the semiclassical approximation
that the highest peak is too high in energy. However this defect is not serious because 
the high-lying states are not important for our analysis.

\section{Effective Potential}
In this appendix, we describe in detail how we encode the inter-site lattice coupling into the 
single-site impurity problem. 
The basic logic is the following. First we write down the energy functional for the $lattice$ problem
in terms of fields labeled by site index $\phi_i$ which couples to some local quantity $\rho_i$, 
then the $local$ partition function is obtained by integrating out all fields 
except the field at origin site $\phi_0$. The long-range order corresponds to some spatial pattern of
$\rho_i$ which generates an extra coupling to local field $\phi_0$.
This extra coupling depends on the long-range order containing information from {\bf other} sites $\rho_i$ $i\neq o$.
We first give a general functional for lattice elastic energy then work out 1-dimensional case explicitly with 
a specific lattice model. Finally we derive the formalism used in our calculation.

\subsection{General Functional of Elastic Energy}
The goal here is to derive the elastic energy in terms of three even-parity MnO$_6$ distortion modes. 
As mentioned in the text, the lattice degree of freedom includes oxygen motion along Mn-O bond $u_i$ and manganese general
displacement $\vec{\delta}_i$. Assuming the spring constant between adjacent Mn-O is $1/K_1$, a general
elastic energy is
\be
E_{lat} &=& \frac{1}{2 K_1} \sum_{i,a} [(\delta^a_i-u^a_i)^2+ (\delta^a_i-u^a_{i-1})^2] \nonumber \\
&+& \frac{1}{2}\sum_{\vec{k},a b} E^{ab}(\vec{k}) \delta^a_{\vec{k}} \delta^b_{-\vec{k}} + 
\frac{1}{2}\sum_{\vec{k},a b} D^{ab}(\vec{k}) u^a_{\vec{k}} u^b_{-\vec{k}} \nonumber \\
&=& \frac{1}{K_1} \sum_{\vec{k},a} [\delta^a_{\vec{k}} \delta^a_{-\vec{k}} + u^a_{\vec{k}} u^a_{-\vec{k}}
- (1+e^{+i k_a}) u^a_{\vec{k}} \delta^a_{-\vec{k}}] \nonumber \\
&+& \frac{1}{2}\sum_{\vec{k},a b} E^{ab}(\vec{k}) \delta^a_{\vec{k}} \delta^b_{-\vec{k}} + 
\frac{1}{2}\sum_{\vec{k},a b} D^{ab}(\vec{k}) u^a_{\vec{k}} u^b_{-\vec{k}}
\ee
where $E^{ab}(\vec{k})$, $D^{ab}(\vec{k})$ represent general harmonic coupling Mn-Mn, O-O displacements, and
$a,b$ sums over $x,y,z$. To get rid of the Mn motions, we use the saddle point approximation 
$\frac{\partial E_{lat}}{\partial (\delta^a_{-\vec{k}})} = 0$ which leads to 
\be
\delta^a_{\vec{k}} =\frac{1}{2 K_1} \sum_b  [ \underline{I} + \underline{E}(\vec{k})/2 )]^{-1}_{ab}  (1+e^{i k_b}) u^b_{\vec{k}}
\ee
and the lattice energy in this approximation is $E_{lat} = \sum_{\vec{k},a b} u^a_{\vec{k}} \tilde{m}^{ab}(\vec{k}) u^b_{-\vec{k}}$ with
\be
\tilde{m}^{ab}(\vec{k}) = \frac{\delta^{ab} }{K_1} - \frac{1}{4 K^2_1}(1+e^{-i k_a}) 
[ \underline{I} + \underline{E}(\vec{k})/2 )]^{-1}_{ab} (1+e^{i k_b}) + \frac{1}{2}D^{ab}(\vec{k})
\ee
Defining strain variables $v_i^a = u_i^a- u_{i-a}^a$, $v_{\vec{k}}^a = u_{\vec{k}}^a (1- e^{-i k_a})$, we express $E_{lat}$ in terms of
$v_{\vec{k}}^a$ which is 
\be E_{lat} = \sum_{\vec{k},a b} v^a_{\vec{k}} m^{ab}(\vec{k}) v^b_{-\vec{k}} \ee
where $m^{ab}(\vec{k}) = \frac{1}{1- e^{-i k_a}} \tilde{m}^{ab}(\vec{k}) \frac{1}{1- e^{i k_b}} $. 
The advantage of expressing $E_{lat}$ in strain variables is that they are closer to the even-parity distortion modes 
defined in Eqn(\ref{eqn:JTmodes}). $m^{ab}(\vec{k})$ allows us to estimate the proximity effect for structural order. 
In particular if we are interested in how $(\pi,\pi,0)$ $Q_x$ order propagates along $z$ direction, then the quantity
to study is $m^{ab}(\pi,\pi,k_z)$. The explicit form of $m^{ab}(\vec{k})$ is model-dependent and 
here we only consider spring constants between adjacent Mn-O ($1/K_1$) and Mn-Mn ($1/K_2$) which are of most importance.

\subsection{1 Dimensional Mn-O Chain}

Now we explicitly work out the local effective potential in the 1-dimensional case. 
The procedure is outlined here. We first adopt the procedure described in the previous subsection to express (with saddle point approximation)
the elastic energy in terms of strain variables in real space $v_i$. Then the effective potential is obtained by
integrating out all $v_i$ except the one at origin $v_0$. 
For 1D Mn-O chain, we drop the index $a,b$ since there is only one direction and the elastic lattice energy is
\begin{eqnarray}
E_{lat} &=&  \frac{1}{2 K_1} \sum_i [(\delta_i-u_i)^2+ (\delta_i-u_{i-1})^2] 
+ \frac{1}{2 K_2} \sum_i (\delta_{i+1}-\delta_i)^2 \nonumber \\
&=& \frac{1}{K_1} \sum_k [u_k u_{-k} + \delta_k \delta_{-k} - u_k
\delta_{-k}(1 + e^{i k})] + \frac{2}{K_2} \sum_k \sin^2(k/2) \delta_k \delta_{-k} 
\end{eqnarray}
For this case, $\frac{E(k)}{2} = \frac{2}{K_2} \sin^2(k/2)$ and $D(\vec{k})=0$. The saddle point approximation 
$\frac{\partial E_{lat}}{\partial \delta_{-k}} = 0$ implies
\begin{equation}
\delta_k = \frac{u_k (1+e^{i k})}{2 + 4 \bar{K} \sin^2(k/2)}
\end{equation}
where $\bar{K} = K_1/K_2$. The effective energy functional $E_{lat}$
(as a function of $u_k$ only) is therefore
\begin{eqnarray}
E_{lat} &=& \frac{2 \bar{K}+1}{K_1} \sum_k \frac{\sin^2(k/2)}{1+2 \bar{K} \sin^2(k/2)} u_k u_{-k}
\nonumber \\ &=&
\frac{2 \bar{K}+1}{4 K_1} \sum_k \frac{1}{1+2 \bar{K} \sin^2(k/2)} v_k v_{-k}
\end{eqnarray}

To see how local strains at different sites couple to one another, we express $E_{lat}$ in the real space $v_i$.
\begin{eqnarray}
E_{lat} &=& \frac{2 \bar{K}+1}{4 K_1} \sum_k \frac{1}{1+2 \bar{K} \sin^2(k/2)} v_k v_{-k}
\nonumber \\ &=&\frac{1}{N} \frac{2 \bar{K}+1}{4 K_1} \sum_{i,j} \sum_k 
\frac{e^{i k (r_i-r_j)}}{1+2 \bar{K} \sin^2(k/2)} v_i v_j \nonumber \\ 
&=& \frac{2 \bar{K}+1}{4 K_1} \sum_{i,j} f(i-j) v_i v_j 
\end{eqnarray}
where $f(i-j) = \frac{1}{N}\sum_k \frac{e^{i k (i-j)}}{1+2 \bar{K} \sin^2(k/2)} 
=\frac{1}{2 \pi} \int_{-\pi}^{\pi} dk \frac{\cos(k n)}{1+2 \bar{K}   \sin^2(k/2)} $
($\sum_k \rightarrow \frac{N}{2 \pi} \int_{-\pi}^{\pi} dk$ for lattice constant $a=1$).
For this simple model, the integral can be done analytically (the most straightforward way may be
changing variable $z = e^{i k}$ and then using the residue theorem!). By
defining $\alpha = 1 + K_2/K_1 $, $E_{lat}$ becomes
\begin{eqnarray}
E_{lat} &=&  \frac{2 \bar{K}+1}{4 K_1} \sqrt{\frac{\alpha-1}{\alpha+1}} 
\sum_{i,j} (\alpha - \sqrt{\alpha^2-1})^{|i-j|} v_i v_j \nonumber \\
&=& \frac{1}{2 K} \sum_{i,j} \epsilon^{|i-j|} v_i v_j 
\end{eqnarray}
where $K = 2 K_1/\sqrt{2 \bar{K}+1}$, and $\epsilon = \alpha - \sqrt{\alpha^2-1} < 1$.
Note that the coupling between local strains is exponentially decay since
$\epsilon^n = e^{-a |n|}$ with $a = -\ln \epsilon$.

Including the the electron-lattice coupling  $\sum_{i} h_i v_i$, the total energy is
\begin{eqnarray}
E = E_{lat} + E_{e-l} = \sum_{ij} A_{ij} v_i v_j + \sum_{i} h_i v_i
\end{eqnarray}
where $h_i$ in this case is the charge density at site $i$.
The effective potential at site 0 is given by integrating out the degrees
of freedom of all other sites $v_1, v_2,...,v_N$, i.e.
\begin{eqnarray*}
\int dv_0 e^{-V_{eff}(v_0)} &=& 
\int dv_0 e^{-A_{00} v_0^2} \, \int dv_1..dv_N 
\exp[-\sum' A_{0i} v_0 v_i - \sum ' A_{ij}v_i v_j - \sum' h_i v_i] \\
&=& \int dv_0 \exp[-(A_{00} - \frac{1}{4} \sum' A_{0i} A^{-1}_{ij} A_{0j} )v_0^2
+ \frac{1}{2} \sum' h_j A^{-1}_{ij} A_{0i} v_0 
+ \frac{1}{4} \sum' h_{i} A^{-1}_{ij} h_{j} ]
\end{eqnarray*}
where $\sum'$ means site 0 is excluded in the summation.

The effective potential is 
\begin{equation}
V_{eff}(v_0) = D v_0^2
- \frac{1}{2} \sum' h_j A^{-1}_{ij} A_{0i} v_0 + \mbox{const}
\end{equation}
where $D = A_{00} - \frac{1}{4} \sum' A_{0i} A^{-1}_{ij} A_{0j}$. We see that the charge density at site $h_i (i \neq 0)$
also contribute to the ``external'' field coupling to $v_0$.



\subsection{3 Dimensional Case}

The 1D result can be easily generalized to the 3D case. For the model we considered, the lattice energy
in $k$ space is
\be
E_{lat} = \frac{2 \bar{K}+1}{4 K_1} \sum_{a=x,y,z} \sum_{k_a}
\frac{1}{1+2 \bar{K} \sin^2(k_a/2)} v_{k_a} v_{-k_a}
\ee
From this expression, we find that in our simple model there is no proximity effect
for $(\pi,\pi,0)$ order of any kind since there is no coupling between different components of $\vec{k}$. 
We also notice that the energy cost is at its minimum when $k_a=\pi$ (staggered order of any kind), therefore
at integer occupancy the system prefer some staggered long-range order since 
the staggered order lowers electronic energy.

Assuming $\epsilon$ is small thus only including the nearest neighbor coupling, the lattice energy in real space is
\be
E_{lat} = \frac{1}{2 K} \sum_{i,a=x,y,z} [ (v_i^a)^2 + 2 \epsilon v_i^a v_{i+a}^a ]
\ee
One can also express $E_{lat}$ in three MnO$_6$ even parity modes $Q$ by the following transformation
\begin{eqnarray*}
\left( \begin{array}{c}
    Q_{i,0}   \\    Q_{i,x} \\ Q_{i,z}
\end{array}  \right) = 
\left( \begin{array}{ccc}
    \frac{1}{\sqrt{3}} & \frac{1}{\sqrt{3}} & \frac{1}{\sqrt{3}}   \\  
    \frac{1}{\sqrt{2}} & -\frac{1}{\sqrt{2}} & 0  \\ 
    -\frac{1}{\sqrt{6}} & -\frac{1}{\sqrt{6}} & \frac{2}{\sqrt{6}} 
\end{array}  \right)
\left( \begin{array}{c}
    v_{i,x}   \\    v_{i,y} \\ v_{i,z}
\end{array}  \right) \equiv U 
\left( \begin{array}{c}
    v_{i,x}   \\    v_{i,y} \\ v_{i,z}
\end{array}  \right)
\end{eqnarray*}
However it is more convenient to work in stain field $v$ until we obtain the local effective potential which will be
expressed in $Q$. 

The electronic source fields $h_i$ are defined as
\begin{eqnarray*}
h_{i,0} &=& \langle e_{ab} c^{\dagger}_{i,a} c_{i,b} \rangle  \\
h_{i,x} &=& \langle \tau^x_{ab} c^{\dagger}_{i,a} c_{i,b} \rangle  \\
h_{i,z} &=& \langle \tau^z_{ab} c^{\dagger}_{i,a} c_{i,b} \rangle  
\end{eqnarray*}
and the local electron-lattice coupling is 
\begin{eqnarray}
E_{JT} = -  h_0 Q_0 -  (h_x Q_x + h_z Q_z) = -(v_x H_x + v_y H_y + v_z H_z )
\end{eqnarray}
with 
\begin{eqnarray}
H_x &=& \frac{1}{\sqrt{3}} h_0 + \frac{1}{\sqrt{2}} h_x - \frac{1}{\sqrt{6}} h_z  \nonumber \\
H_y &=& \frac{1}{\sqrt{3}} h_0 - \frac{1}{\sqrt{2}} h_x - \frac{1}{\sqrt{6}} h_z  \nonumber \\
H_z &=& \frac{1}{\sqrt{3}} h_0 + \frac{2}{\sqrt{6}} h_z 
\label{eqn:Hfield}
\end{eqnarray}

Following the procedure for 1-dimensional case, 
the effective potential at the origin site is therefore 
\begin{eqnarray}
V_{eff} &=& D (v_x^2 + v_y^2 + v_z^2) + \frac{1}{2} \vec{E} \cdot \vec{v} \\
&=& D (Q_0^2 + Q_x^2 + Q_z^2) + \frac{\epsilon}{2} \vec{F} \cdot \vec{Q}
\label{eqn:selfh}
\end{eqnarray}
with $E_{x(y,z)} = \sum' H_{j_{x(y,z)}, x(y,z)} A^{-1}_{i_{x(y,z)}, j_{x(y,z)} } A_{0, i_{x(y,z)}} $
and $\vec{F} = U \vec{E}$. $i_{x(y,z)}$ labels the sites along $x (y,z)$  axis.
We call $\vec{F}$ effective external field. Keeping only the linear term in $\epsilon$, we have
\begin{eqnarray}
D &=& \frac{1}{2K} (1-\epsilon^2/4) \sim \frac{1}{2K} \\
E_x &=&  H_{+\hat{x},x} + H_{-\hat{x},x}  \nonumber \\
E_y &=&  H_{+\hat{y},y} + H_{-\hat{y},y}  \nonumber \\
E_z &=&  H_{+\hat{z},z} + H_{-\hat{z},z}   \\
\end{eqnarray}
and $\vec{F} = \left( \frac{1}{\sqrt{3}} (E_x+E_y+E_z),  
\frac{1}{\sqrt{2}} (E_x-E_y), \frac{1}{\sqrt{6}} (-E_x-E_y+2E_z) \right)$.


\bibliographystyle{physrev}
\bibliography{prb_draft_BulkManganite}

\end{document}